\documentclass[12pt]{article}

\usepackage{amsfonts}
\usepackage{amsmath}
\usepackage{amssymb}
\usepackage[dvips]{graphicx}
\usepackage{comment}

\newcommand{\ba}{\begin{eqnarray}}
\newcommand{\ea}{\end{eqnarray}}
\newcommand{\be}{\begin{equation}}
\newcommand{\ee}{\end{equation}}
\newcommand{\nn}{\nonumber}
\newcommand{\hJ}{\hat{J}}
\newcommand{\hL}{\hat{L}}
\newcommand{\vY}{\vec{Y}}
\newcommand{\vW}{\vec{W}}

\newcommand{\sqz}{\mathbf{z}}

\newcommand{\cW}{\mathcal{W}}
\newcommand{\rrangle}{\rangle\!\rangle}
\newcommand{\llangle}{\langle\!\langle}

\def\a{\alpha}

\def\d{\delta}

\def\k{\kappa}
\def\l{\lambda}

\newcommand{\nt}{\nonumber\\}

\newcommand{\bZ}{{\bf Z}}

\newcommand{\bra}[1]{ \langle {#1} | }
\newcommand{\ket}[1]{ | {#1} \rangle }

\newcommand{\ban}{\begin{eqnarray*}}
\newcommand{\ean}{\end{eqnarray*}}

\allowdisplaybreaks[1]

\begin{document}

\begin{titlepage}

\begin{flushright}
UT-12-22
\end{flushright}

\vskip 12mm

\begin{center}
{\Large Virasoro constraint for Nekrasov instanton partition function}
\vskip 2cm
{\large Shoichi Kanno, Yutaka Matsuo and Hong Zhang}
\vskip 2cm
{\it Department of Physics, The University of Tokyo}\\
{\it Hongo 7-3-1, Bunkyo-ku, Tokyo 113-0033, Japan}
\end{center}
\vfill
\begin{abstract}
We show that Nekrasov instanton partition function for 
$SU(N)$ gauge theories satisfies
recursion relations in the form of  $U(1)+$Virasoro
constraints when $\beta=1$. 
The constraints give a direct support for 
AGT conjecture for general quiver gauge theories.
\end{abstract}
\vfill
\end{titlepage}

\setcounter{footnote}{0}

\section{Introduction and Summary}
Some years ago, Nekrasov and his collaborators \cite{r:Nekrasov}
found an exact form of the instanton partition functions of 
$\mathcal{N}=2$ supersymmetric gauge theories
in omega background with two deformation parameters
$\epsilon_1,\epsilon_2$.  It became a milestone in the understanding
of supersymmetric gauge theories and their connection
with 2D integrable system and initialized the later developments (see for example, \cite{r:AGT,Nekrasov:2009rc}).

In this paper, along the line of such developments,
we claim that there exist simple recursion
formulae for Nekrasov's instanton partition function
for $SU(N)$ gauge theories 
with $\beta=-\epsilon_2/\epsilon_1=1$ written
in the following form:
\ba
\sum_{\vY',\vW'} (\hJ_n)_{\vY,\vW} ^{\vY',\vW'}  Z_{\vY',\vW'}=0
\,,\qquad
\sum_{\vY',\vW'}(\hL_n)_{\vY,\vW} ^{\vY',\vW'} Z_{\vY',\vW'}=0\,,
\label{cJL}
\ea
where $\hJ_n$ and $\hL_n$ ($n\in \mathbf{Z}$) 
are infinite dimensional
matrices which satisfy Virasoro + U(1) current algebra without central extension,
\ba
[\hJ_n,\hJ_m]=0,\quad
[\hL_n, \hJ_m]=-m \hJ_{n+m},\quad
[\hL_n, \hL_m]=(n-m)\hL_{n+m}\,.\label{u1vir}
\ea
The indices $\vec Y, \vec W$ are collections of $N$ Young diagrams,
for example, $\vec Y=(Y_1,\cdots,Y_N)$.
$Z_{\vec Y,\vec W}$ is a part of the instanton partition function
consisting of the contribution of vector and bifundamental multiplets.
Precise forms of $\hJ_n, \hL_n, Z_{\vec Y,\vec W}$  will be given in the text.
We conjecture that the relations are part of more general
$\cW_{1+\infty}$ algebra where 
(\ref{u1vir}) become the subalgebra.

The constraint equations give a direct support to
($SU(N)$ generalization of) AGT
conjecture \cite{r:AGT} which suggests the partition functions of $\mathcal{N}=2$ theories
equal to the conformal block functions of Liouville (Toda) field theory.
As we will review in next section, the instanton 
partition function for  $SU(N)\times\cdots\times SU(N)$ 
quiver gauge theory is written out of $Z$ in (\ref{cJL}), 
and (\ref{u1vir}) implies the existence of conformal Ward identity
in the conformal block functions.  One remarkable feature is that
the proof is not restricted by the number of boxes of Young diagrams
but holds in all orders analytically.

The constraint of the form (\ref{cJL}) appeared in various contexts
in string theory.  A famous example is the matrix model which describes
two dimensional gravity (Virasoro constraint \cite{r:Virasoro,r:IM1}
and $\cW_{1+\infty}$ constraint \cite{r:W, r:IM2}).
Given the intimate relation between the matrix model
and AGT conjecture\cite{Dijkgraaf:2009pc}, 
the existence of such relation is quite natural.

We organize the paper as follows.  In section \ref{s:Nekrasov},
we define the Nekrasov function $Z_{\vec Y,\vec W}$ in (\ref{cJL}).
It is a building block of instanton partition function for linear quiver gauge theories.
In section \ref{s:CFT}, we propose a formula which represents
$Z_{\vec Y\vec W}$ as a 3-point function of conformal field theory.   
While it is written in terms of free fermions,
the direct calculation of the correlation function is nontrivial  since
there is room for inserting screening operators.  Instead of directly
computing the correlator, we show the conformal Ward identity written
in the form (\ref{u1vir}).  Finally in section \ref{s:proof}, 
we give a direct proof of such recursion formula in terms of
Nekrasov function. Since the proof is technical and lengthy,
we write some explicit computation in the appendix.

\section{Nekrasov partition function}\label{s:Nekrasov}
\begin{figure}[bpt]
\begin{center}
\includegraphics[scale=0.7]{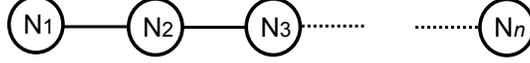}
\end{center}
\caption{Linear quiver}
\label{f:quiver}
\end{figure}

\begin{figure}[bpt]
\begin{center}
\includegraphics[scale=0.5]{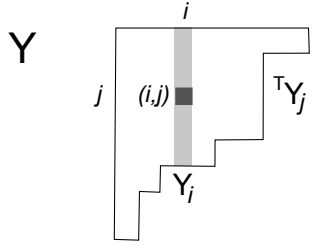}
\end{center}
\caption{Young diagram}
\label{f:coord}
\end{figure}

In this paper, we focus on the linear quiver gauge theories
with gauge group $SU(N_1)\times\cdots \times SU(N_n)$ (Figure \ref{f:quiver}).
For this case, Nekrasov partition function
is written in the form of matrix multiplication \cite{r:AGT,Alba:2009ya},
\ba\label{e:Nek}
Z^\mathrm{Nek}= \sum_{\vec Y^{(1)},\cdots,\vec Y^{(n)}}
q_i^{|\vec Y^{(i)}|} 
\bar{V}_{\vec Y^{(1)}} \cdot Z_{\vec Y^{(1)} \vec Y^{(2)}}\cdots
Z_{\vec Y^{(n-1)}\vec Y^{(n)}} \cdot V_{\vec Y^{(n)}}\,.
\ea
Here $q_i=e^{2\pi i \tau_i}$ describes the coupling constant $\tau_i$ for
$i^\mathrm{th}$ gauge group $SU(N_i)$.
Sets of Young tables $\vec Y^{(i)}=(Y^{(i)}_1,\cdots,Y^{(i)}_{N_i})$ are 
used to label the contribution from fixed points of the localization technique
and $|\vec Y^{(i)}|=\sum_{p=1}^{N_i} |Y^{(i)}_p|$ is the sum of the
number of boxes for each Young diagram. Each ``matrix" $Z$ or ``vector"
$V,\bar{V}$ contains the information of vacuum expectation value
for vector multiplet $\vec a^{(i)}$
associated with each gauge group $SU(N_i)$, the mass for 
$SU(N_i)$-$SU(N_{i+1})$ bifundamental multiplet $\mu^{(i)}$,
and the mass for fundamental (anti-fundamental) multiplet $\vec\lambda$,
($\vec\lambda'$).
They are explicitly written in terms of a function $Z$,
\ba
Z_{\vec Y^{(i)} \vec Y^{(i+1)}}&=& Z(\vec a^{(i)},\vY^{(i)}; \vec a^{(i+1)}, \vY^{(i+1)};\mu^{(i)}),\\
\bar V_{\vY^{(1)}}&=&  Z(\vec\lambda, \vec\emptyset; \vec a^{(1)}, \vY^{(1)};\mu^{(0)}),\\
V_{\vec Y^{(n)}}&=& Z(\vec a^{(n)},\vY^{(n)}; \vec \lambda', \vec\emptyset;\mu^{(n)}),
\ea
where $\vec\emptyset$ is a set of null Young diagrams $(\emptyset,\cdots,\emptyset)$
and 
\ba
Z(\vec a,\vY; \vec b, \vW;\mu)&=& 
\sqz_\mathrm{vect}(\vec a,\vec Y)\bar\sqz_\mathrm{vect}(\vec b, \vec W)
z_\mathrm{bifund}(\vec a,\vec Y; \vec b,\vec W;\mu) \,,\\
\sqz_\mathrm{vect}(\vec a,\vec Y)
&=& \prod_{p,q=1}^N
G_{Y_p,Y_q}(a_p-a_q)^{-1} \,,\\
\bar\sqz_\mathrm{vect}(\vec a,\vec Y)
&=& \prod_{p,q=1}^N (-1)^{|Y_q|}
G_{Y_q,Y_p}(1-\beta-a_p+a_q)^{-1}\,,\\
z_\mathrm{bifund}(\vec a,\vec Y; \vec b,\vec W;\mu)
&=&  \prod_{p,q=1}^N(-1)^{|W_q|}
G_{Y_p,W_q}(a_p-b_q-\mu)G_{W_q,Y_p}(1-\beta-a_p+b_q+\mu)\,,\\
G_{A,B}(x)&:=& \prod_{(i,j)\in A} (x+\beta(({}^TA)_j-i+1)+((B)_i-j))\,.
\ea
$\sqz_\mathrm{vect}$ and $\bar\sqz_\mathrm{vect}$ 
are related to the usual factor for the vector multiplet
as $z_\mathrm{vect}(\vec a,\vec Y)= \sqz_\mathrm{vect}(\vec a,\vec Y)
\cdot\bar\sqz_\mathrm{vect}(\vec a,\vec Y)$.
${}^\mathtt{T}\! A$ is the transpose of a Young table $A$, 
$(i,j)$ is the coordinate of a box in the Young diagram $A$ and $(A)_i$ (resp. ${}^\mathtt{T}\! (A)_j$)
represents the height of $i^\mathrm{th}$ colomn (resp. 
the length of $j^\mathrm{th}$ row).  See Figure \ref{f:coord} for the illustration.

For technical reasons, we restrict our analysis to $\beta=1$ case
and derive the $U(1)$+Virasoro constraint
for $Z(\vec a,\vY; \vec b, \vW;\mu)$ with $N_i=N_{i+1}$.
We will argue that similar constraints exist also
for $V_{\vec Y}, \bar V_{\vec Y}$.
They can be interpreted as a proof of $U(1)$+conformal symmetry
in Nekrasov function.

\section{2D CFT}\label{s:CFT}
\subsection{A conjectured relation}
The $SU(N)$ generalization of
AGT conjecture implies that the 
partition function (\ref{e:Nek}) can be written as the
conformal block of 
$n+3$ point function of $SU(N)$ Toda field theory
\cite{Wyllard:2009hg,Mironov:2009by}
where the Hilbert space $\mathcal{H}$ is described by
chiral $W_n$ algebra with $U(1)$ factor.

\begin{figure}[bpt]
\begin{center}
\includegraphics[scale=0.6]{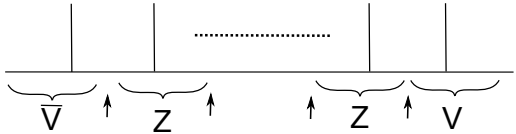}
\end{center}
\caption{Conformal block and correspondence with Nekrasov factor}
\label{f:block}
\end{figure}

We write the conformal block in Figure \ref{f:block}.
It can be
reduced to the multiplication of three point functions
by inserting a complete basis of
the Hilbert space at the intermediate channel.
In Figure \ref{f:block}, insertion points of such operators
are depicted by arrows. In $W_n$ +$U(1)$ system, the basis
of the Hilbert space is labeled by $N$ Young tables $\vec Y$.
Then it may be possible to choose such basis such that
the factor $Z_{\vec Y,\vec W}$ in the previous section may be rewritten as
$Z_{\vec Y,\vec W}\sim \langle \vec Y|V(1)|\vec W\rangle$
with some vertex operator $V$.
The existence of such basis was formally claimed in \cite{Alba:2010qc,Fateev:2011hq}
for general $\beta$ in terms of Jack polynomial, but the explicit form was not given
except for some simple examples.

An exceptional case occurs when $\beta=1$ and the system is described
by $N$ pairs of free fermions.  In this case, there is a reasonable guess
on the explicit form of $|\vec Y,\vec a\rangle$ \cite{Belavin:2011js,Kanno:2011qv}
as a product of  Schur polynomials, 
namely $|\vec Y\rangle \sim \prod_{p=1}^N \chi_{Y^{(p)}}$.
(See also \cite{Estienne:2011qk} for a similar analysis.)
In the following, we will provide more precise definition of such states
in the Hilbert space of free fermion
including the background charges.  The formula
we would like to establish is 
\ba\label{conj}
Z(-\vec a,\vY; -\vec b, \vW;\mu)=\llangle \vY, \vec a +\nu\vec{e}|V_\k(1) |\vW, \vec b+(\nu-\mu)\vec{e}\rrangle\,,
\ea
where $V_\k$ is a vertex operator and $\vec{e}=(1,1,\dots,1)$.
The parameter $\nu$ is arbitrary.
The vertex operator must has a special form of a momentum $(\k,0,\cdots,0)$
to satisfy the Virasoro constraint, where $\k$ is determined by
the $U(1)$ charge conservation,
\ba\label{consrv}
\sum_i (a_i+\nu)=-\k+\sum_j (b_j+\nu-\mu).    
\ea
The number of parameters in gauge theory  and CFT  is matched
up to the irrelevant parameter $\nu$.
Precise definitions of the basis $|\vec W, \vec\lambda\rangle$, 
the vertex operator $V_\kappa$
and the inner product $\llangle \cdot|\cdot\rrangle$ are explained in the following subsections.

\subsection{Free fermion and vertex operator}
We start from the definition of fermions,
\ba
\bar\psi^{(p)}(z) =\sum_{n\in \bZ} \bar\psi^{(p)}_n z^{-n-\lambda_p
-1} \,,\;\;\; \psi^{(p)}(z)  =\sum_{n\in \bZ} \psi^{(p)}_n z^{-n+\lambda_p}\,,
\quad p=1,\cdots, N,\quad
z\in \mathbf{C}
\ea
with anti-commutation relation, $\{ \bar\psi^{(p)}_n,\psi^{(q)}_m \}=\d_{p,q}\d_{n+m,0}$.
We note that there are extra parameters $\vec\l\in \mathbf{R}^N$ which 
represent the shift of the usual mode expansion of fermion.
We define the vacuum as,
$\ket{\vec{\l}}=\otimes_{p=1}^N |\lambda^{(p)}\rangle$,
\ba
\bar\psi^{(p)}_n\ket{\vec{\l}}=\psi^{(p)}_m\ket{\vec{\l}}=0 \;\;\; (n \geq 0, m>0),\quad
\vec\l=(\l^{(1)},\cdots,\l^{(N)})\,.
\ea
The parameters $\vec\lambda$ represent the fermion sea levels.
Similarly, the bra vacuum $\bra{\vec{\l}}=\otimes_{p=1}^N \bra{\lambda^{(p)}}$ is defined by
\ba
\bra{\vec{\l}}\bar\psi^{(p)}_n=\bra{\vec{\l}}\psi^{(p)}_m=0 \;\;\; (n < 0, m\leq0),\quad
\vec\l=(\l^{(1)},\cdots,\l^{(N)})\,.
\ea
In formula (\ref{conj}), the bra state has different sea level (say $\vec\mu$) in general.
In such cases, we need redefine fermion mode expansion
as $ \psi^{(p)}(z)  =\sum_{n\in \bZ} \psi^{(p)}_n z^{-n+\lambda_p}
=\sum_{n\in \bZ} \tilde\psi_n^{(p)} z^{-n+\mu_p}$
and define the bra vacuum in terms of $\tilde\psi$.
The Hermitian conjugate is defined as $(\ket{\vec{\l}})^{\dagger}=\bra{\vec{\l}}$ 
and $\psi_n^{\dagger}=\bar\psi_{-n}$. This is consistent
with the shift of label by the change of vacuum.

With this preparation, the basis used in (\ref{conj})
is (after translated in the fermion basis),
\ba\label{fbasis}
|\vec{Y},\vec\l \rangle &=& \otimes_{p=1}^N\left(
\bar\psi^{(p)}_{-\bar r_1^{(p)}} \bar\psi^{(p)}_{-\bar r_2^{(p)}}\cdots 
\bar\psi^{(p)}_{-\bar r_{s_1}^{(p)}} |\lambda^{(p)},
-s_1\rangle\right) ,\quad
|\lambda^{(p)},-s_1\rangle = \psi_{-s_1+1}^{(p)}\cdots \psi_{-1}^{(p)}
 \psi_0^{(p)}|\lambda^{(p)}\rangle\\
 \label{fbasis2}
 &=&(-1)^{|\vec Y|}   \otimes_{p=1}^N\left(
\psi^{(p)}_{-\bar s_1^{(p)}} \psi^{(p)}_{-\bar s_2^{(p)}}\cdots 
\psi^{(p)}_{-\bar s_{r_1}^{(p)}} |\lambda^{(p)},
r_p\rangle\right) ,\quad
|\lambda^{(p)},r_1\rangle = \bar\psi_{-r_1}^{(p)}\cdots \bar\psi_{-1}^{(p)}
 |\lambda^{(p)}\rangle \\
\bra{\vec{Y},\vec{\l}}&=&(\ket{\vec{Y},\vec{\l}})^\dagger
\ea
Here we represent a Young diagram $Y_p$
by the number of each row $r_\sigma^{(p)}=({}^\mathtt{T}Y_p)_\sigma$
or the number of each columns $s_\sigma^{(p)}=(Y_p)_\sigma$. 
The parameters with bar are $\bar r_{\sigma}^{(p)}=
r_{\sigma}^{(p)}-\sigma+1$
and $\bar s_\sigma^{(p)}=s_\sigma^{(p)}-\sigma$.  
These states give a natural basis of the Hilbert space
with fixed fermion number.  
By construction,
they are orthonormal  $\langle \vec Y,\vec a| \vec W, \vec b\rangle=
\delta_{\vec Y, \vec W}\delta_{\vec a,\vec b}$\,.

We define the vertex operator $V_{\kappa}$ in (\ref{conj}) by 
standard bozonization technique. We write,
\ba
\psi^{(p)}(z)=:e^{-\phi_p(z)}:,\quad
\bar\psi^{(p)}(z)=:e^{\phi_p(z)}:\,,
\ea
with
\ba
\phi_p(z)=x_p+ a_0\log z -\sum_{n\neq 0}\frac{a^{(p)}_n}{n} z^{-n}\,,\quad
[a^{(p)}_n, a^{(q)}_m]=n\delta_{p,q}\delta_{n+m,0},\ 
[x_p, a^{(q)}_0]=\delta_{p,q}\,.
\ea
The vacuum and the fermionic basis (\ref{fbasis}) is written in a form,
\ba
|\vec\lambda\rangle = \lim_{z\rightarrow 0} :e^{-\sum_p\lambda_p\phi_p(z)}:|\vec{0}\rangle,
\quad
|\vec{Y},\vec\l \rangle =\prod_p \chi_{Y^{(p)}}(a^{(p)}_{-n})|\vec\lambda\rangle\,.
\ea
Here $ \chi_{Y^{(p)}}(a^{(p)}_{-n})$ is Schur polynomial expressed
in terms of power sum $\mathbf{p}_n=\sum_i (x_i)^n$ and each $\mathbf{p}_n$
is replaced by $a^{(p)}_{-n}$.  While the second expression is not used in the following,
it is this expression that appeared in the literature \cite{Fateev:2011hq,Belavin:2011js,Kanno:2011qv}.
The vertex operator in (\ref{conj}) is written as,
\ba
V_{\vec\kappa}(z) =:e^{\sum_p \kappa_p \phi_p(z)}:\,.
\ea

Here we have to be careful in the definition of the inner product (\ref{conj}).
If we interpret it as the correlation function for free fields,
the momentum for each fermion pair should be separately
conserved, namely,
\ba
\langle \vec Y,\vec a+\nu \vec e|V_{\vec\kappa}(1) |\vec W, \vec b+(\nu-\mu)\vec e\rangle\propto
\delta_{\vec a,\vec\kappa+\vec b-\mu \vec e}\,.
\ea
On the other hand, in the Nekrasov formula, there is no such constraints.
So we have to interpret the inner product not as that of free fields but
the conformal block of $W_n$ algebra +$U(1)$ current algebra
as in the literature.  The difference between the two is that
one may insert the screening operators to recover the
conservation of momentum.\footnote{ An example of such interacting 
system is $c=1$ Liouville theory \cite{Schomerus:2003vv}.}
While it is not explicitly written in (\ref{conj}),
the insertion of screening operators
is implicitly assumed.  In general it gives a generalized Selberg
integral of Schur functions \cite{Itoyama:2010ki, Mironov:2010pi, Zhang:2011au}
when a set of Young tables is empty, namely $\vec W=\vec\emptyset$.
For this case, the integration path of the screening currents may be taken as paths
connecting  $0$ to $1$.

In our case where both $\vec Y, \vec W$ are not empty, the definition
of such integration is even more tricky.  
We will not attempt to do this in this paper but use 
(\ref{conj}) as a formal expression to derive the recursion formulae
that Nekrasov function should obey.
The proof of the identity does not use the definition (\ref{conj})
but the properties of Nekrasov function alone.

Another point we have to pay attention to is the momentum of the vertex operator.
As we mentioned in the previous subsection, we need to take it as a special form
$\vec\kappa=(\kappa,0,\cdots,0)$.
This is required by the closure of conformal Ward identity 
for $W$ algebra \cite{Wyllard:2009hg,Mironov:2009by}.
We will come back to this issue later.

\subsection{$\cW_{1+\infty}$ algebra}
For $\beta=1$ case, $W_N$ +$U(1)$ current algebra
is enhanced to $\cW_{1+\infty}$
algebra \footnote{Some explicit relations are given in  \cite{Kanno:2011qv}.},
which is a quantization of the algebra
of higher differential operators.  
For a differential operator $z^nD^m$
($D=z\frac{\partial}{\partial z}$), we define a generator
$\cW(z^n e^{xD}):=\sum_{m=0}^\infty
\frac{x^m}{m!} \cW(z^n D^m)$ as,
\ba\label{e:fermion}
\cW(z^ne^{xD})&=&\frac{1}{2\pi i}\oint_{z=0}dz\sum_{p=1}^N z^n :\bar\psi^{(p)}(z)e^{xD} 
\psi^{(p)}(z):-\sum_{p=1}^N \frac{e^{\l_p x}-1}{e^x-1}\delta_{n,0}\\
&=&\sum_{p=1}^N\sum_{\ell \in \mathbf{Z}}e^{x (\ell+\lambda_p)} :\bar\psi^{(p)}_{\ell+n}\psi^{(p)}_{-\ell}:
-\sum_{p=1}^N \frac{e^{\l_p x}-1}{e^x-1}\delta_{n,0}.
\ea
From our definition of the Hermitian conjugation, we see $W(z^nD^m)^\dagger=W(z^{-n}(D-n)^m)$.
Their commutation relation is written  as,
\ba\label{Winf}
[\cW(z^ne^{xD}) , \cW(z^me^{yD})]=(e^{mx}-e^{ny})\cW(z^{n+m} e^{(x+y)D})-C \frac{e^{mx}-e^{ny}}{e^{x+y}-1}\delta_{n+m,0}
\,,
\ea
with $C=N$. The realization (\ref{e:fermion}) gives a unitary representation of 
$\cW_{1+\infty}$ \cite{Awata:1994tf}.  The $U(1)$ current and Virasoro
operators are embedded in $\cW_{1+\infty}$ as,
\ba
J_n&=&\cW(z^n)=\sum_{p=1}^N \sum_{m\in \mathbf{Z}} :\bar\psi^{(p)}_{n+m}\psi^{(p)}_{-m}:
-\delta_{n0}\sum_{p=1}^N\lambda_p,
\\
 L_n&=&-\cW(z^nD)-\frac{n+1}{2}\cW(z^n)
 =-\sum_{p=1}^N \sum_{m\in \mathbf{Z}}\frac{ n+2m+2\lambda_p+1}{2}
  :\bar\psi^{(p)}_{n+m}\psi^{(p)}_{-m}:+\frac{\delta_{n0}}{2}\sum_{p=1}^N\lambda_p^2,
  \label{Ln}
\ea
which satisfy 
\ba
[J_n,J_m]=N\delta_{n+m,0},\quad
[L_n, J_m]=-m J_{n+m},\quad
[L_n, L_m]=(n-m)L_{n+m}+\frac{N}{12}(n^3-n)\delta_{n+m,0}\,.\label{u1vir2}
\ea
While the bosonized version of $\cW_{1+\infty}$ generators are given in
a closed form \cite{Kanno:2011qv}, they are in general highly nonlinear.
The exceptions are $U(1)$ and Virasoro generators which have the standard form,
\ba
J_n=\frac{1}{2\pi i}\oint_{z=0}dz \sum_{p=1}^{N} z^n \partial \phi_p(z),\qquad
L_n=\frac{1}{2\pi i}\oint_{z=0}dz \sum_{p=1}^{N} z^{n+1} \frac{1}{2}:(\partial \phi_p(z))^2:.
\ea
In the following, we treat the inner product (\ref{conj}) as the
conformal block of $\cW_{1+\infty}$ algebra.  The use of $\cW_{1+\infty}$
instead of $W_N+U(1)$ has definite 
merit in the simplicity of the expression (\ref{e:fermion})
and the algebra (\ref{Winf}) in a closed form.

The screening operators of $\cW_{1+\infty}$ mentioned in the previous subsection
are written as,
\ba
S_{pq}=\int_{\mathcal{C}} d\zeta \bar\psi^{(p)}(\zeta)\psi^{(q)}(\zeta)\,,
\ea
with $p\neq q$.  It can be easily established that it commute with all the generators
of $\cW_{1+\infty}$ algebra as long as the integration contour $\mathcal{C}$
is appropriately chosen \cite{Dotsenko:1984nm}.  We assume these operators are implicitly
inserted in (\ref{conj}).  We used a notation $\llangle\cdot|\cdot \rrangle$ to implement this idea,
\ba
\llangle \vec Y,\vec \lambda|V_\k(1)|\vec W, \vec\mu\rrangle
:=\langle \vec Y,\vec \lambda|V_\k(1)S_{p_1q_1}S_{p_2 q_2}\cdots|\vec W, \vec\mu\rangle\,.
\ea
Here the right hand side is the inner product of the free fields.
The insertions of screening operators change the U(1) charge for each boson in the form
$(c_1,\cdots, c_N)$ with $\sum_{i=1}^N c_i=0$.
The conservation of momentum for each $\phi_p$
can be broken but only their sum is conserved, namely 
we have (\ref{consrv}).  

An intriguing feature of the fermion basis  (\ref{fbasis}) is that the action of $\cW_{1+\infty}$
generators is written neatly.  In particular, they are simultaneous
eigenstates of all the commuting generators of $\cW_{1+\infty}$,
\ba
\cW(e^{xD})|\vec{Y},\vec\l \rangle &=& \Delta(\vec Y, \vec \l, x)|\vec{Y},\vec\l \rangle,
\label{eigen}\\
\Delta(\vec Y, \vec \l, x)&=&\sum_{p=1}^N \left(
\sum_{\sigma_p=1}^{s_p} \left(
e^{-x r^{(p)}_{\sigma_p}}-1
\right)e^{x(\sigma_p-1+\l_p)}
\right)-\sum_{p=1}^N \frac{e^{\l_p x}-1}{e^x-1}
\label{eigen1}
\\
&=&\sum_{p=1}^N \left(
\sum_{\sigma_p=1}^{r_p} \left(1-
e^{x s^{(p)}_{\sigma_p}}
\right)e^{x(-\sigma_p+\l_p)}
\right)-\sum_{p=1}^N \frac{e^{\l_p x}-1}{e^x-1}\,.
\label{eigen2}
\ea
In particular,
\ba
J_0|\vec{Y},\vec\l \rangle=-\sum_{p=1}^N\lambda_p |\vec{Y},\vec\l \rangle,\quad
L_0|\vec{Y},\vec\l \rangle=\sum_{p=1}^N\left(|Y_p|+\frac{\lambda_p^2}{2} \right)|\vec{Y},\vec\l \rangle\,.
\ea

\subsection{Construction of the constraints}
Now we arrive at the position to explain how to construct the 
recursion relation of the form (\ref{cJL}).
The conjectured relation (\ref{conj}), while it is not
completely well-defined, gives a good hint.
We use the following trivial identity \footnote{
A nontrivial example was examined in \cite{Zhang:2011au}.},
\ba
0&=&
\llangle \vY, \vec a +\nu \vec{e}|\cW(z^n D^p) V_\k(1) |\vW, \vec b+(\nu-\mu)\vec{e}\rrangle\nn\\
&&- \llangle \vY, \vec a +\nu \vec{e}|V_\k(1) \cW(z^n D^p )  |\vW, \vec b+(\nu-\mu)\vec{e}\rrangle\nn\\
&&-\llangle \vY, \vec a +\nu \vec{e}|[\cW(z^n D^p ) , V_\k(1)] |\vW, \vec b+(\nu-\mu)\vec{e}\rrangle\nn\\
&=&\sum_{\vec Y', \vec W'} \hat \cW (z^n D^p )_{\vec Y,\vec W}^{\vec Y', \vec W'}
\llangle \vY', \vec a +\nu \vec{e}|V_\k(1) |\vW', \vec b+(\nu-\mu)\vec{e}\rrangle
\,.\label{e:WI}
\ea
Here the first two lines can be evaluated by action of
$\cW(z^n D^p)$ on the bra and ket basis.
The third line is given by the commutator with the vertex.
As we see these are written as a linear combination of
inner product and can be written in the fourth line.
The insertion of screening charges does not play any role since
they commute with $\cW_{1+\infty}$ generators.
The coefficients of the recursion relations
$\hat\cW (z^n D^p )_{\vec Y,\vec W}^{\vec Y', \vec W'}$
satisfy the
$\cW_{1+\infty}$ algebra since they are the difference between
the action of $\cW(z^n D^p)$ on the bra and vertex$+$ket states.
The central charges cancel between the two terms.

If the eq.(\ref{conj}) holds, the Nekrasov function should also satisfy the relation,
namely,
\ba\label{e:W}
\sum_{\vec Y', \vec W'} \hat \cW (z^n D^p )_{\vec Y,\vec W}^{\vec Y', \vec W'}
Z(-\vec a,\vY'; -\vec b, \vW';\mu)=0.
\ea
This is what we would like to establish in the following.
 
Actually we meet a technical problem in the computation  of $\cW(z^n D^p)$ with
$p\geq 2$.  Since their bosonic realization is highly nonlinear, 
the commutator with the vertex operator becomes 
messy.  So in this paper, we limit ourselves to focus on safer
$U(1)$+Virasoro part ($p=0,1$).
We also note that we do not need to derive all the identity of the form
(\ref{e:WI}).  Since they form a noncommutative algebra,
proving identity of the form (\ref{e:WI}) for 
 $\cW(z^{\pm 1})$ and $\cW(z^{\pm n}D)$ $n=1,2$ 
  (i.e. $J_{\pm 1},L_{\pm 1}, L_{\pm 2}$) will generate
 all other constraints.  For example, $[\hat L_1, \hat J_1]=-\hat J_2$, 
 $[\hat L_2, \hat L_1]=\hat L_3$ and so on.
 
 In the following we evaluate the action
 of $\cW_{1+\infty}$ on the basis and the vertex.


\subsubsection*{Action on bra and ket basis}
\begin{figure}[bpt]
\begin{center}
\includegraphics[scale=0.6]{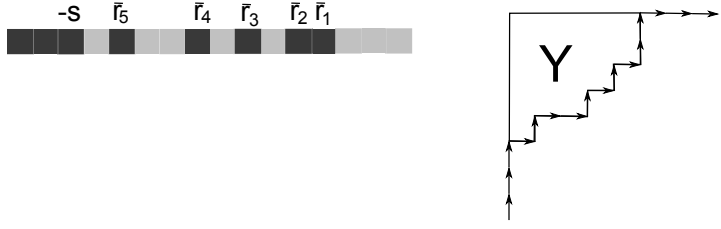}
\end{center}
\caption{Young diagram and fermion state}
\label{f:Maya}
\end{figure}

In order to evaluate
the action of $W(z^n e^{xD})$ ($n\neq 0$) on $|\vec{Y},\vec\l \rangle$,
a graphic representation (Maya diagram) of $|\vec{Y},\vec\l \rangle$ \cite{r:Sato} is useful.
For the simplicity of argument, we take $N=1$ and remove the
the index $p$ in (\ref{fbasis},\ref{fbasis2}).  We take the first expression (\ref{fbasis})
and rewrite it as,
\ba
|Y,\lambda\rangle =\bar\psi_{-\bar{r}_1}\bar\psi_{-\bar{r}_2}\cdots\bar\psi_{-\bar{r}_s}
\bar\psi_{s}\bar\psi_{s+1}\cdots\bar\psi_{L}|-L,\lambda\rangle\,.
\ea
and take $L\rightarrow \infty$ limit.  From this representation, we associate a 
Young diagram $Y$ with a semi-inifinite 
sequence of integers $S_Y=\{ \bar{r}_1,\bar{r}_2,\cdots, \bar{r}_s, -s,-s-1,\cdots \}$.
We prepare an infinite strip of boxes with integer label and fill the boxes with
the integer in $S_Y$ (Figure \ref{f:Maya} left).
It represents the occupation of fermion in each level.
To understand the correspondence with the Young diagram $Y$,
we associate each black box with vertical up arrow and white box
with horizontal right arrow.  We connect these arrows for each box
from the left on $S_Y$.  Then the Young diagram shows up
in the up/left corner (Figure \ref{f:Maya} right).  
The generator $\cW(z^n e^{xD})=\sum_\ell e^{x(\ell+\lambda)}:\bar\psi_{\ell+n}\psi_{-\ell}:$
flips one black box at $\ell$ to  white and one white box at $-\ell-n$ to black
(if wrong color was filled at each place, it vanishes).  It amounts to flipping
vertical arrow by horizontal one and vice versa.  By analyzing the effect of such flipping,
the action of $\cW(z^n e^{xD})$ on $|Y,\lambda\rangle$ can be summarized as,
\begin{figure}[bpt]
\begin{center}
\includegraphics[scale=0.6]{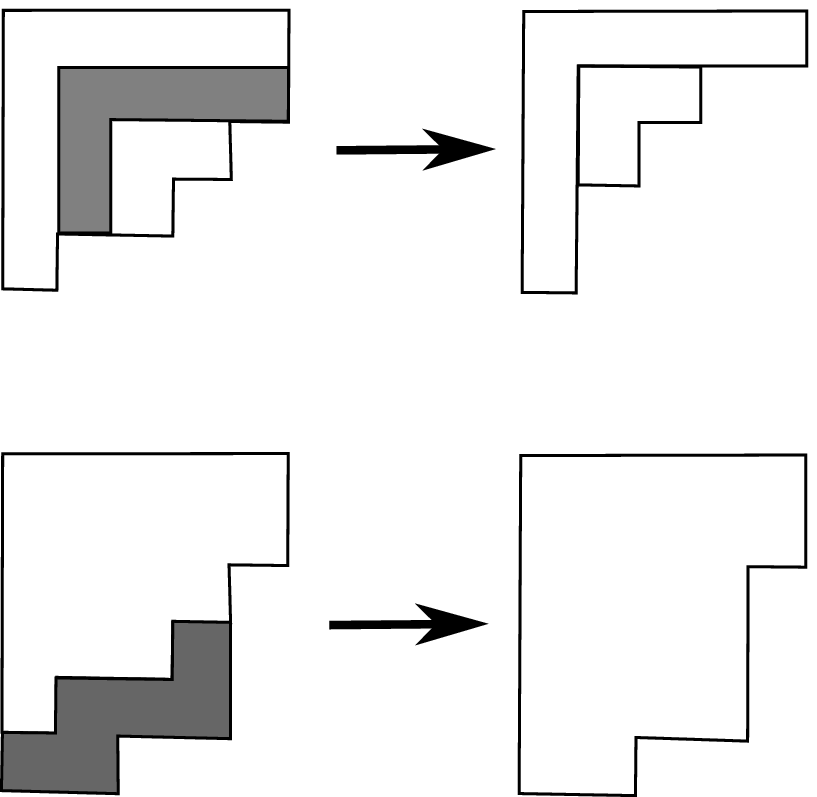}
\end{center}
\caption{Action of $\cW$ on $|Y\rangle$}
\label{f:add}
\end{figure}
\begin{itemize}
\item For $n>0$ it erases a hook of length $n$ and multiply $(-1)^{v(h)-1} e^{x(\ell+\lambda)}$
where $v(h)$ is the height of the hook.  (Figure \ref{f:add} up)
If there are some hooks of length $n$, we sum over all such possibilities.
\item For $n<0$ it adds a strip of length $|n|$ and multiply $(-1)^{v(h)-1} e^{x(\ell+\lambda)}$
where $v(h)$ is the height of the strip.  (Figure \ref{f:add} down)
As in $n>0$ case, if there are some possibility, we need to add them.
\end{itemize}
As we explained, in practice we need evaluate only $n=\pm1, n=\pm2$ cases.
The action of $\cW(z^n D^m)$ is much simplified
and the explicit form is given in section \ref{s:proof}.

\subsubsection*{Commutator with the vertex}
The vertex operator $V_\kappa(1)$ is the operator version of state
$|\vec \kappa\rangle$.
As we mentioned, it is restricted to be of the form,
$\vec \kappa=(\kappa,0,\cdots,0)$
and the vertex
is expressed as $e^{\kappa\phi_1(1)}$.\footnote{
Actually the vector like $(0,\cdots, 0,\kappa,0,\cdots,0)$ works as well.
We need only one component to be non-vanishing.}
This is a restricted set of vacuum where we have only two independent
states at level one, namely $W(z^{-1})|\vec \mu\rangle$ and $W(z^{-1}D)|\vec \kappa\rangle$.
All other states are related to the second one as, $W(z^{-1}D^p)|\vec \kappa\rangle
=\kappa^{p-1}W(z^{-1}D)|\vec \kappa\rangle$ for $p\geq 1$.
This is the  level one degenerate state condition
for simple vertex \cite{Wyllard:2009hg,Mironov:2009by} for $\cW_{1+\infty}$.
In the next section, we show that the vertex to have this form
is necessary to have even  $U(1)$ and Virasoro constraints.

The derivation of the commutation relation between the
vertex and $U(1)$ and Virasoro generator are straightforward since it is a primary field,
\ba\label{JLV}
[J_n,V_{\kappa}(1)]=\kappa V(1),\quad
[L_n, V_{\kappa}(1)]=\frac{\kappa^2(n+1)}{2} V_\kappa(1)+\partial V(1)\,.
\ea
On the other hand, the operator $W(z^n D^m)$ with $m\geq 2$
is written in terms of boson as $W(z^n D^m)\sim (\partial \phi)^{m+1}$
and the commutation relation with the vertex is not written in a compact form.
The exceptional case is $\kappa=\pm 1$ where the vertex operator can be identified
with the fermion,
\ba
[\cW(z^n D^m), \bar\psi(\zeta)]= \zeta^n D_\zeta^m \bar\psi(\zeta),\qquad
[\cW(z^n D^m), \psi(\zeta)]= -(\zeta^n D_\zeta^m)^\dagger \psi(\zeta)\,.
\ea
Except for such cases, the evaluation of recursion formula becomes complicated.
While we spent some time to solve this problem, we could not manage
to write it in a closed form.  Because of this technical issue, we will not
analyze $\cW(z^n D^m)$ with $m\geq 2$.

\section{Proof of $U(1)$/Virasoro constraints}\label{s:proof}
As we mentioned, the derivations of recursion formulae for $J_{\pm 1}$,
$L_{\pm 1}, L_{\pm 2}$ will be enough to prove (\ref{cJL}).
We will explicitly show them one by one in this section.

Here, we give a few remarks.  
\begin{itemize}
\item As a corollary of  (\ref{cJL}),  
we can obtain the recursion formulae for $V_{\vec Y}, \bar V_{\vec Y}$
in (\ref{e:Nek}) automatically if the reader follow the proof in the following.
These can be identified
with the Nekrasov partition function 
$SU(N)$ gauge theory with $2N$ fandamental matters.
All we need to do is to
restrict $\vec W=\vec \emptyset$ and restrict the constraints to $J_1,L_1,L_2$.
They are proved by observing, for example, 
$\cW(z^n D^m)|\vec\emptyset, \vec b+(\nu-\mu)\vec e\rangle=0$
for $n>0$ in the arguments below.
By taking commutator among constraints, 
it gives rise to a family of constraints of the form,
\ba
 (\hat J_{n})_{\vec Y}^{\vec Y'} V_{\vec Y}=(\hat L_{n})_{\vec Y}^{\vec Y'} V_{\vec Y}=0,
\ea
with $n> 0$. This can be regarded as the $U(1)$+Virasoro constraints
for fundamental+vector multiplets.

\item In the computation below, we will analyze the recursion relation
when the rank of $\vec Y$ and $\vec W$ in (\ref{conj}) can be
different, namely $\vec Y=(Y_1,\cdots, Y_N)$ and $\vec W=(W_1,\cdots, W_M)$
without requiring $N=M$.
In CFT, such possibility is difficult to interpret since it implies 
we have different number of fermions on the bra and ket states.  On the other hand, in the 
context of linear quiver, it shows up when the ranks of gauge groups
are different. While Nekrasov formula exists for such cases $N\neq M$,
the interpretation in terms of CFT 
has been a focus of the literature \cite{r:KMS, r:DP}.
Our analysis in the following implies $N=M$ to keep the constraints.
This is natural from the viewpoint of CFT.   This seems to support the claim 
in \cite{r:KMS} that AGT type conjecture holds only
for $SU(N)\times\cdots\times SU(N)$ type quiver.
\end{itemize}

\subsection{$U(1)$}
\begin{figure}[bpt]
\begin{center}
\includegraphics[scale=0.4]{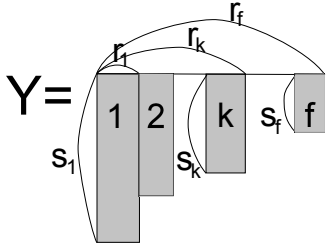}
\end{center}
\caption{Rectangle decomposition of Young diagram}
\label{f:Young}
\end{figure}

\begin{figure}[bpt]
\begin{center}
\includegraphics[scale=0.4]{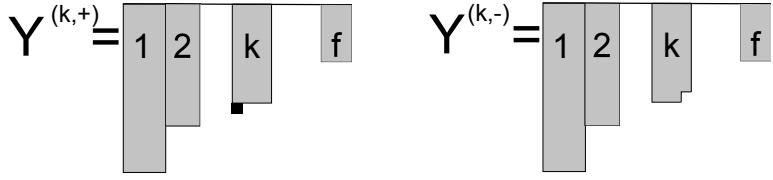}
\end{center}
\caption{Adding (subtracting) a box to/from Young diagram}
\label{f:Young+-}
\end{figure}

From the explanation in the previous section, the action of $\cW(z^{\pm 1} D^m)$
on the basis $|\vec Y, \lambda\rangle$ (or $\langle \vec Y, \lambda|$)
is obtained by adding or subtracting one box on one of the Young diagram in 
$\vec Y$ with appropriate coefficient.  
This can be more explicitly expressed
by representing  each Young diagram $Y_p$ as a set of rectangles
(Figure \ref{f:Young}).
We denote the Young diagram in the figure as,
$
Y=[(r_1,s_1),\cdots,(r_f,s_f)]
$.
($r_1<r_2<\cdots<r_f$, $s_1>s_2>\cdots>s_r$).
We write $p^\mathrm{th}$ diagram in $\vY,\vW$ as,
\ba
Y_p=[(r^{(p)}_1,s^{(p)}_1),\cdots,(r^{(p)}_{f^{(p)}},s^{(p)}_{f^{(p)}})]\,,\quad
W_p=[(t^{(p)}_1,u^{(p)}_1),\cdots,(t^{(p)}_{\bar f^{(p)}},u^{(p)}_{\bar f^{(p)}})]\,.
\ea
The addition or the subtraction of a box is expressed on (from) which rectangle
the box is added (subtracted).
We denote $Y^{(k,+)}$ (resp. $Y^{(k,-)}$) as the diagram by adding 
(resp. removing) a box in $k^\mathrm{th}$ rectangle of $Y$ (Figure \ref{f:Young+-}).

With this notation, the first three lines in (\ref{e:WI}) are 
evaluated as,
\ba
&&\langle \vec Y,\vec a+\nu\vec e|J_1=\sum_{p=1}^N \sum_{k=1}^{f_p+1}\langle \vec Y^{(k,+),p},\vec a+\nu\vec e|, \label{Ja}\\
&& [J_1, V_\k(1)]=\k V(1)\,,\label{Jb}\\
&& J_1 |\vec W,\vec b+(\nu-\mu)\vec e\rangle = 
\sum_{q=1}^M \sum_{l=1}^{\bar{f}_q}|\vec W^{(l,-),q},\vec b+(\nu-\mu)\vec e\rangle \label{Jc}.
\ea
Here $\vY^{(k,+),p}=(Y_1,\cdots Y_p^{(k,+)},\cdots Y_N)$ and
$\vW^{(k,-),p}=(W_1,\cdots W_p^{(k,-)},\cdots W_M)$.
From these expressions, the coefficient $\hat J_1$ in the last line of (\ref{e:WI}) is written as
\ba
(\hat J_1)_{\vec Y,\vec W}^{\vec Y',\vec W'}=
\left\{
\begin{array}{ll}
-\k \quad & \vec Y=\vec Y',\, \vec W=\vec W'\\
1 \quad & \vec Y^{(k,+),p}=\vec Y', \vec W=\vec W' \\
-1 \quad & \vec W^{(k,-),p}=\vec W',\  \vec Y=\vec Y'\\
0 \quad &\mbox{otherwise}
\end{array}
\right.\,.
\ea 
Similarly, for $W(z^{-1})=J_{-1}$,
\ba
(\hat J_{-1})_{\vec Y,\vec W}^{\vec Y',\vec W'}=
\left\{
\begin{array}{ll}
-\k\quad & \vec Y=\vec Y',\, \vec W=\vec W'\\
1 \quad & \vec Y^{(k,-),p}=\vec Y', \vec W=\vec W''\\
-1 \quad & \vec W^{(k,+),p}=\vec W',\  \vec Y=\vec Y'\\
0 \quad &\mbox{otherwise}
\end{array}
\right.\,.
\ea 


We put these explicit forms to  (\ref{e:W})
and prove the identity.  For this purpose, we need evaluate
the quantity,
\ba
Q(\vY',\vW'; \vY,\vW) \equiv
\frac{Z(-\vec a,\vY'; -\vec b, \vW';\mu)}{Z(-\vec a,\vY; -\vec b, \vW;\mu)}\,,
\ea
with $\beta=1$.
The constraint for $J_{1}$ is written as,
\ba
&&\sum_{p=1}^{N} \sum_{k=1}^{f_{p+1}}Q(\vY^{(k,+),p},\vW; \vY,\vW)
= \k+\sum_{q=1}^M \sum_{k=1}^{\bar f_p} Q(\vY,\vW^{(k,-),p}; \vY,\vW)
\label{J1a}\,.
\ea
Since the proof for $J_{-1}$ is completely parallel, we focus to give
the explicit computation for $J_1$.

We evaluate the change by the addition and subtraction of a box in 
Young diagrams in Nekrasov formula.\footnote{
It seems rather straightforward to compute it for general $\beta$
but the variation of factors with two different Young diagrams in Nekrasov formula
is difficult to evaluate.  For the computation, the lemma 4 in \cite{Zhang:2011au}
is essential but it holds only when $\beta=1$.\label{lemma4}}
After a lengthy computation (see appendix \ref{s:prq+-} for detail), 
two $Q$'s in (\ref{J1a}) are evaluated as,
\ba
Q(\vY^{(k,+),p},\vW; \vY,\vW)
& = &
(-1)^{N+M-1}
\prod_{q =1}^{N}
 \frac
{\prod_{l=1}^{f_q} A_{(p)k}-B_{(q)l} } 
{\prod_{(q)l\neq (p)k}^{f_q+1} A_{(p)k}-A_{(q)l}}
\times
\prod_{q=1}^{M}
 \frac{\prod_{l=1}^{f_q+1} A_{(p)k}-C_{(q)l}}
{\prod_{l=1}^{f_q} A_{(p)k}-D_{(q)l} } 
\, ,
\label{Q+}\\
Q(\vY,\vW^{(k,-),p}; \vY,\vW)
&=&
(-1)^{N+M} 
\prod_{q =1}^{N}
 \frac
{\prod_{l=1}^{f_q} D_{(p)k}-B_{(q)l} } 
{\prod_{l=1}^{f_q+1} D_{(p)k}-A_{(q)l}}
\times
\prod_{q=1}^{M}
 \frac{\prod_{l=1}^{f_q+1} D_{(p)k}-C_{(q)l}}
{\prod_{(q)l\neq (p)k}^{f_q} D_{(p)k}-D_{(q)l} } 
\, ,
\label{Q-}
\ea
where
\begin{equation}
\begin{array}{ll}
A_{(p)k}=a_p +\nu+s^{(p)}_k - r^{(p)}_{k-1}, \qquad  &1\leq p\leq N, \qquad 1\leq k \leq f_p+1 \, ,\\
B_{(p)k}=a_p +\nu + s^{(p)}_k - r^{(p)}_{k} , \qquad  &1\leq p\leq N,\qquad 1\leq k \leq f_p \, , \\
C_{(p)k}=b_p+\nu-\mu + u^{(p)}_k - t^{(p)}_{k-1} , \qquad &1\leq p\leq M,\qquad 1\leq k \leq \tilde{f}_p+1 \, ,\\
D_{(p)k}=b_p+\nu-\mu + u^{(p)}_k - t^{(p)}_{k} , \qquad &1\leq p\leq M ,\qquad 1\leq k \leq \tilde{f}_p \, .
\end{array}\label{ABCD}
\end{equation}
We further rewrite
\ba 
&& A_{(1)k},A_{(2)k}, \dots, A_{(n)k},
D_{(N+1)k},D_{(N+2)k}, \dots, D_{(N+M)k}
 \equiv x_I, \\
&&B_{(1)k},B_{(2)k}, \dots, B_{(N)k},
C_{(N+1)k},C_{(N+2)k}, \dots, C_{(N+M)k}
 \equiv -y_J .
\ea
The index $I$ goes from $1$ to
$\sum_{p=1}^N  (f_p +1 )+ \sum_{q=1}^M \bar{f}_q
=N+\sum_{p=1}^{N} f_p +\sum_{q=1}^M \bar{f}_q \equiv \mathcal N $, 
whereas $J$ goes from $1$ to
$M+\sum_{p=1}^{N} f_p +\sum_{q=1}^M  \bar{f}_q \equiv \mathcal M$.
Two terms in Eq.(\ref{J1a}) that contain $Q$ are rewritten in a compact form,
\ba
&&\sum_{p=1}^{N} \sum_{k=1}^{f_{p+1}}Q(\vY^{(k,+),p},\vW; \vY,\vW)-
\sum_{q=1}^M \sum_{k=1}^{\bar f_p} Q(\vY,\vW^{(k,-),p}; \vY,\vW) \nt
&&=(-1)^{N+M-1}
\sum_{I=1}^{\mathcal N}
\frac{\prod_{J=1}^{\mathcal M}(x_I +y_J)}{\prod_{J \neq I}^{\mathcal N}(x_I 
-x_J)}\,.\label{Q-Q}
\ea
Then one may use an identity (see the appendix \ref{s:prcomb} for a proof),
\ba\label{xy}
\sum_{I=1}^{\mathcal N}
\frac{\prod_{J=1}^{\mathcal M}(x_I +y_J)}{\prod_{J \neq I}^{\mathcal N}(x_I 
-x_J)}=\mbox{Coefficient of $\zeta^{M-N+1}$ of}\quad
\frac{\prod_{J=1}^\mathcal{M} (\zeta+y_J)}{\prod_{I=1}^\mathcal{N} (\zeta-x_I)}  \, .
\ea
In particular, for $\mathcal{N}=\mathcal{M}$ (i.e. $N=M$)
the right hand side of (\ref{Q-Q}) gives,
\ba
-\sum_{I=1}^\mathcal{N} (x_I+y_I) =-\sum_{p=1}^{N}(a_p-b_p+\mu)=\k \,.
\ea
In the second equality, we use the charge conservation which is derived from the Ward identity for $J_0$. 
This proves the constraint of $J_1$ for $N=M$.  At the same time,
it implies the constraint holds only when $N=M$.

\subsection{Virasoro}
Next, let us consider Virasoro constraint. The analogs of eqs. (\ref{Ja})-(\ref{Jc})
for $L_1$  are
\ba
&&\langle \vec Y,\vec a+\nu\vec e|L_1=-\sum_{p=1}^N \sum_{k=1}^{f_p+1}(a_p+\nu+s_k^{(p)}-r_{k-1}^{(p)})\langle \vec Y^{(k,+),p},\vec a+\nu\vec e|, \label{La}\\
&& [L_1, V_\k(1)]=( \partial_\zeta+\k^2) V(\zeta)|_{\zeta=1} \,,\label{Lb}\\
&& L_1 |\vec W,\vec b+(\nu-\mu)\vec e\rangle = -\sum_{q=1}^M \sum_{l=1}^{\bar{f}_q}(b_q+\nu-\mu+u_l^{(q)}-t_{l}^{(q)}) |\vec W^{(l,-),q},\vec b+(\nu-\mu)\vec e\rangle \label{Lc}.
\ea
The coefficients in  (\ref{La}),(\ref{Lc}) are more complicated than $U(1)$
case because the Virasoro generators have the derivative $D$ (\ref{Ln}).
(\ref{Lb}) comes from (\ref{JLV}).

The matrix elements of $(\hL_1)_{\vY,\vW} ^{\vY',\vW'}$ are given by
\ba
(\hL_1)_{\vY,\vW} ^{\vY',\vW'} =\left\{
\begin{array}{ll}
-\frac{1}{2}|\vec{a}+\nu \vec{e}|^2+\frac{1}{2}|\vec{b}+(\nu-\mu)\vec{e}|^2-\frac{1}{2}\k^2-|\vec Y|+|\vec W| 
& \vec Y=\vec Y',\, \vec W=\vec W'\\
-(a_p+\nu+s_k^{(p)}-r_{k-1}^{(p)}) &\vY^{(k,+),p}=\vec Y',\, \vec W=\vec W'\\
b_q+\nu-\mu+u_l^{(q)}-t_{l}^{(q)}  &\vec Y=\vec Y',\,\vW^{(k,-),p}=\vec W'\\
0 & \mbox{otherwise}
\end{array} \right.
\ea
where we use 
\ba
\llangle \vY, \vec a|V_\mu(\zeta) |\vW, \vec b\rrangle\ \propto \zeta^{ \frac{1}{2}|\vec{a}|^2
-\frac{1}{2}|\vec{b}|^2-\frac{1}{2}\mu^2+|\vec Y|-|\vec W|}
\ea
to evaluate the term which contains the derivative of vertex operator. This is derived from the Ward identity for $L_0$.
Similarly, those of $(\hL_{-1})_{\vY,\vW} ^{\vY',\vW'}$ are given by
\ba
(\hL_{-1})_{\vY,\vW} ^{\vY',\vW'} =\left\{
\begin{array}{ll}
-\frac{1}{2}|\vec{a}+\nu \vec{e}|^2+\frac{1}{2}|\vec{b}+(\nu-\mu)\vec{e}|^2+\frac{1}{2}\k^2-|\vec Y|+|\vec W| 
& \vec Y=\vec Y',\, \vec W=\vec W'\\
-(a_p+\nu+s_k^{(p)}-r_{k}^{(p)}) &\vY^{(k,-),p}=\vec Y',\, \vec W=\vec W'\\
b_q+\nu-\mu+u_l^{(q)}-t_{l-1}^{(q)}  &\vec Y=\vec Y',\,\vW^{(k,+),p}=\vec W'\\
0 & \mbox{otherwise}
\end{array} \right.
\ea
Ward identity for $L_1$ is rewritten as
\ba
&&-\sum_{p=1}^{N} \sum_{k=1}^{f_{p+1}}(a_p+\nu+s_k^{(p)}-r_{k-1}^{(p)})Q(\vY^{(k,+),p},\vW; \vY,\vW) \nt
&&~~~~~~~~~~~+\sum_{q=1}^M \sum_{k=1}^{\bar f_p} (b_q+\nu-\mu+u_l^{(q)}-t_{l}^{(q)})Q(\vY,\vW^{(k,-),p}; \vY,\vW)\nt
&&~~~~~~~~~~~~~~~~~~~
=( \frac{1}{2}|\vec{a}+\nu \vec{e}|^2-\frac{1}{2}|\vec{b}+(\nu-\mu)\vec{e}|^2+\frac{1}{2}\k^2+|\vec Y|-|\vec W| )
\label{L1a}\, .
\ea

We see that the coefficient in front of $Q$s in (\ref{L1a}) are $A_{(p)k}$ and $D_{(p)k}$
in (\ref{ABCD}). 
Therefore, the left hand side of (\ref{L1a}) can be written as
\ba
(-1)^{N+M}\sum_{I=1}^{\mathcal N}x_I
\frac{\prod_{J=1}^{\mathcal M}(x_I +y_J)}{\prod_{J \neq I}^{\mathcal N}(x_I -x_J)}
\label{QL}. 
\ea
For $N=M$,
\ba
\sum_{I=1}^{\mathcal N}x_I
\frac{\prod_{J=1}^{\mathcal N}(x_I +y_J)}{\prod_{J \neq I}^{\mathcal N}(x_I -x_J)}
=\sum_{I<J} (x_I x_J+y_I y_J)+\sum_I x_I^2 +(\sum_I x_I)(\sum_J y_J)\,.
\label{LQ2}
\ea
After some calculation (see appendix \ref{a:Virasoro}), 
we see that (\ref{LQ2}) is exactly equal to the right hand side of (\ref{L1a}).

Our proof for $L_{\pm2}$ is almost the same as $L_{\pm1}$. The difference is that $L_{\pm2}$ increase or decrease two
connected boxes when they act on the bra or ket states. There are two ways to add (subtract) the two connected boxes on
the corner of each rectangles. One way is to add vertically lined boxes(we name it $Y^{(k,+2E)}$) and the other is to 
add horizontal lined boxes($Y^{(k,+2H)}$).
After a tedious calculation(see appendix \ref{a:Virasoro}), the part which comes from a variation of Young diagram can be expressed as 
\ba
\frac{1}{2}\sum_I^{2\mathcal N} \tilde{x}_{I} 
\frac{\prod_{J}^{2\mathcal M} \tilde{x}_{I}+\tilde{y}_{J}}
{\prod_{J \neq I}^{2\mathcal N} \tilde{x}_{I}-\tilde{x}_{J}}  \, ,\label{L2id}
\ea
where
\ba
\tilde{x}_{I}&=&\left\{
\begin{array}{ll}
x_I \;\;&(I=1,\cdots,\mathcal N) \\
x_{I-\mathcal {N}} -1\;\;&(I=\mathcal{N}+1,\cdots,2 \mathcal{N})
\end{array} \right.\\
\tilde{y}_{J}&=&\left\{
\begin{array}{ll}
y_J \;\;&(J=1,\cdots,\mathcal M) \\
y_{J-\mathcal {M}} +1\;\;&(J=\mathcal{M}+1,\cdots,2 \mathcal{M})
\end{array} \right.\,.
\ea

When the width of some rectangle or 
the difference of height between two adjoining rectangles
is one, we can not add 
two boxes at that location so some terms are lacked to express the variation as (\ref{L2id}).
But, in such a case, the corresponding terms in (\ref{L2id}) become zero (see appendix \ref{a:Virasoro}). 
If we get rid of all the meaningless zero terms from the summation, it reduces to the right formula. 
In other words, by adding suitable zero terms, we can get (\ref{L2id})
for any Young diagrams with arbitrary shape.

\section{Discussion}

In this paper, we give a direct proof
that  Nekrasov partition
function satisfies Virasoro and U(1) constraints
which strongly support AGT conjecture.  

As we mentioned in the text, there are some direct
extensions of the analysis made here.
One is to extend the constraint
to $\cW_{1+\infty}$ algebra.  For that purpose, it will be
sufficient to give the recursion formula for $\cW(D^2)$ since
the commutation with $J_{\pm1}$ gives all other generators.
While the action of $\cW(D^2)$ to fermion basis is diagonal
(\ref{eigen}--\ref{eigen2}), the commutator with $V_\kappa$
is nontrivial.
This is related to the 
fact that the vertex operator does not transform in covariant
way in $\cW_{1+\infty}$.  Since
the existence of such constraint proves AGT conjecture
for $\beta=1$, this is an important challenge. We hope to have
technical improvement to answer this question in the near future.

Another issue is to consider general $\beta$.
In our case, this is again due to a technical difficulty that the
variation of Nekrasov formula is much harder to obtain
(see footnote \ref{lemma4}).
This may, however, be a more
profound issue.  For $\beta=1$ case, the symmetry of the system
is identified with $\cW_{1+\infty}$ algebra.  It is known
that the unitary representation of $\cW_{1+\infty}$ algebra
is limited to free fermion system, namely $\beta=1$ case.
For general $\beta$, we need some kind of deformed version of
$\cW_{1+\infty}$ algebra.  Since $\cW_{1+\infty}$ algebra plays 
essential role in various places in theoretical physics \cite{r:W, r:IM2, r:Winf},
the deformation of $\cW_{1+\infty}$ is certainly a challenging problem. 
Recently in a mathematical literature \cite{r:SV}, the action of
deformed version of $\cW_{1+\infty}$ on the fixed points was  given
for general $\beta$.  There will be certainly some hope to work in this direction.

\subsection*{Acknowledgement}
Two of the authors (SK and YM) would like to thank
the hospitality of colleagues in Saclay where part of the
work was carried out. We would like
to thank J. Bourgine, T. Kimura, I. Kostov, V. Paquier, 
S. Ribault, C. Rim, R. Santachiara, D. Serban, S. Shiba, Y. Tachikawa for various
discussions, comments and encouragements.  This work is partially
supported by Sakura project (collaboration program
between France and Japan) by MEXT Japan.
S.K. is partially supported by Grant-in-Aid (\#23-10372)
for JSPS Fellows.
YM is partially supported by Grand-in-Aid
(KAKENHI \#20540253) from MEXT Japan.
HZ is partially supported by Global COE Program, 
the Physical Sciences Frontier, MEXT, Japan.

\appendix
\small
\section{Proof of Eqs.(\ref{Q+}, \ref{Q-})}\label{s:prq+-}
We introduce some notations,
\ba
G_{A,B}(x) &=& \prod\limits_{(i,j) \in A} \Big(   x + \beta (({}^T\!A)_j - i) + ((B)_i - j) + \beta \Big) \; ,\\
g_{Y_p, W_q}(a_p-b_q- \mu) &=& G_{Y_p, W_q}(a_p-b_q- \mu)
G_{W_q,Y_p }(-a_p+b_q+ \mu +1 -\beta) \; ,
\ea
where in the first line, $A$ and $B$ are Young tables.
The left hand side of  eqs.(\ref{Q+}, \ref{Q-}) is written as,
\ba
Q(\vY^{(k,+),p},\vW; \vY,\vW)
&=&\prod_{q=1}^{M} \frac{g_{Y^{(k,+)}_p, W_q}(b_q-a_p- \mu)}
{g_{Y_p, W_q}(b_q-a_p- \mu)}
\times
\prod_{q\neq p}^{N} \frac
{g_{Y_p, Y_q}(a_q-a_p)}{g_{Y^{(k,+)}_p, Y_q}(a_q-a_p)}
\times
\frac
{G_{Y_p, Y_p}(0)}{G_{Y^{(k,+)}_p, Y^{(k,+)}_p}(0)}\,,
\label{Q+a}
\\
Q(\vY,\vW^{(k,-),p}; \vY,\vW)
&=&\prod_{p=1}^{N} \frac{g_{Y_p, W^{(k,-)}_q}(b_q-a_p- \mu)}
{g_{Y_p, W_q}(b_q-a_p- \mu)}
\times
\prod_{p\neq q}^{M} \frac
{g_{W_p, W_q}(b_q-b_p)}{g_{W_p, W^{(k,-)}_q}(b_q-b_p)}
\times
\frac
{G_{W_p, W_p}(0)}{G_{W^{(k,-)}_p, W^{(k,-)}_p}(0)}\,.
\label{Q-a}
\ea
The evaluation of the last term is relatively straightforward.
For example,
\begin{equation}
\begin{split}
\frac{G_{Y,Y}(0)}{G_{Y^{(k,+)},Y^{(k,+)}}(0)} 
&= 
\prod_{i=1}^{r_{k-1}} \frac{r_{k-1} +(Y)_i -i-s_k}{r_{k-1} +(Y)_i -i+1-s_k}
\times
\prod_{j=1}^{s_k} \frac{({}^T\! Y)_j + s_k -r_{k-1}-j }{({}^T\! Y)_j + s_k -r_{k-1}-j +1 } 
\\&= 
\prod_{l=k}^{f_i} \frac{r_{k-1} +s_l -r_l -s_k}{r_{k-1} +s_{l+1} -r_l -s_k}
\times
\prod_{l=1}^{k-1} \frac{r_{k-1} +s_l -r_l -s_k}{r_{k-1} +s_{l} -r_{l-1} -s_k}
\\&= 
\frac{\prod_{l=1}^{f_i} r_{k-1} +s_l -r_l -s_k}
{\prod_{l\neq k}^{f_i +1}r_{k-1} +s_{l} -r_{l-1} -s_k}\,.
\end{split}
\end{equation}
Direct evaluation of the first two terms in (\ref{Q+a},\ref{Q-a})
turns out to be rather nontrivial.  We need to use the lemma 4 of
\cite{Zhang:2011au} where it was proved,
\be
g_{A, B}(x)=(-1)^{|A|+|B|}[N_2-x]_A[x-N_1]_B
\times 
\prod_{j=1}^{N_{1}}\prod_{i=1}^{N_{2}}
 \frac{ x +1 + ({}^T\! A)_{j} + (B)_{i}-i -j }
{ x +1 -i -j } \, ,
\ee
with $[x]_A=\prod_{(i,j)\in A} (x-i+j)$ and $N_1, N_2$ be arbitrary  integers
which are larger than heights and widths of Young diagrams $A, B$.  
It can be
shown that the right hand side does not depend on $N_1, N_2$.

We evaluate the ratio of factors by using this formula,
\ba
\frac{g_{A^{(k,+)}, B}(x)}{g_{A, B}(x)}
&=&-\frac{[N_2-x]_{A^{(k,+)}}}{[N_2-x]_A}
\times 
\prod_{j=1}^{N_{1}}\prod_{i=1}^{N_{2}}
 \frac { x +1 + ({}^T\! A^{(k,+)})_{j} + (B)_{i}-i -j }
{ x +1 + ({}^T\! A)_{j} + (B)_{i}-i -j }
 \nn\\
&=&(x+r^A_{k-1}-s^A_{k}-r^B_{f_B})
\times
\prod_{l=1}^{f_B}
 \frac{x +r^A_{k-1} +s^B_l -r^B_{l-1} -s^A_k}
{x +r^A_{k-1} +s^B_l -r^B_l -s^A_k }       \nn
\\
&=&
 \frac{\prod_{l=1}^{f_B+1} x +r^A_{k-1} +s^B_l -r^B_{l-1} -s^A_k}
{\prod_{l=1}^{f_B} x +r^A_{k-1} +s^B_l -r^B_l -s^A_k } \,,  \\        
\frac{g_{A, B^{(k,-)}}(x)}{g_{A, B}(x)}
&=&-
 \frac
{\prod_{l=1}^{f_A}x +r^A_{l} +s^B_k -r^B_{k} -s^A_l } 
{\prod_{l=1}^{f_A+1}x +r^A_{l-1} +s^B_k -r^B_{k} -s^A_{l}} \,.
\ea
Here we have made use of the following relations
\ba
&&\frac{[ x ]_{A^{(k,+)}}}{[ x ]_A}
=x -r_{k-1} +s_k\,,
\quad
\frac{[ x ]_{B^{(k,-)}}}{[ x ]_B}
=\frac{1}{x-r_{k} +s_k}\,,\\
&& \begin{split}
&\prod_{j=1}^{N_{1}}\prod_{i=1}^{N_{2}}
 \frac 
{ x +1 + ({}^T\! A^{(k,+)})_{j} + (B)_{i}-i -j }{ x +1 + ({}^T\! A)_{j} + (B)_{i}-i -j }
=
\prod_{i=1}^{N_{2}}
 \frac
{ x +1 + r^A_{k-1} + (B)_{i}-i -(s^A_k +1) +1 }{ x +1 + r^A_{k-1} + (B)_{i}-i -(s^A_k +1) }
\\
&~~~~~~~~~~=
 \frac
{x +r^A_{k-1}  -s^A_k-r^B_{f_B}}{x +r^A_{k-1}-s^A_k -N_2}
\times
\prod_{l=1}^{f_B}
 \frac
{x +r^A_{k-1} +s^B_l -r^B_{l-1} -s^A_k}{x +r^A_{k-1} +s^B_l -r^B_l -s^A_k } \,,
\end{split}\\
&& \begin{split}
&\prod_{j=1}^{N_{1}}\prod_{i=1}^{N_{2}}
 \frac
{ x +1 + ({}^T\! A)_{j} + (B^{(k,-)})_{i}-(i +j) }{ x +1 + ({}^T\! A)_{j} + B_{i}-(i +j) }
=
 \frac
{x -r^B_{k}+s^B_k -N_1}{x -r^B_{k}+s^B_k -s_1^A}
\times
\prod_{l=1}^{f_A}
 \frac{x +r^A_{l} +s^B_k -r^B_{k} -s^A_l }
{x +r^A_{l} +s^B_k -r^B_{k} -s^A_{l+1}} \,.
\end{split}
\ea
After combining these factors we obtain,
(here we rename $W_q \equiv Y_{N+q}$, $b_q -\mu \equiv a_{N+q}$, 
$u^{(q)} \equiv s^{(N+q)}$, and $t^{(q)} \equiv r^{(N+q)}$)
\ba
\prod_{q=1}^{M} \frac{g_{Y^{(k,+)}_p, W_q}(b_q-a_p- \mu)}
{g_{Y_p, W_q}(b_q-a_p- \mu)}
&=&
\prod_{q=N+1}^{N+M} \frac{g_{Y^{(k,+)}_p, Y_q}(a_q-a_p)}
{g_{Y_p, Y_q}(a_q-a_p)} 
\nn\\
&=&(-1)^M\prod_{q=N+1}^{N+M}
\bigg \{
 \frac{\prod_{l=1}^{f_q+1}a_p-a_q -r^{(p)}_{k-1} -s^{(q)}_l +r^{(q)}_{l-1} +s^{(p)}_k}
{\prod_{l=1}^{f_q} a_p-a_q -r^{(p)}_{k-1} -s^{(q)}_l +r^{(q)}_l +s^{(p)}_k } 
\bigg \}      \,, \\
\prod_{q \neq p}^{N} \frac
{g_{Y_p, Y_q}(a_q-a_p)} {g_{Y^{(k,+)}_p, Y_q}(a_q-a_p)}
&=&(-1)^{N-1}\prod_{q \neq p}^{N}
\bigg \{
 \frac
{\prod_{l=1}^{f_q} \a_p-\a_q -r^{(p)}_{k-1} -s^{(q)}_l +r^{(q)}_l +s^{(p)}_k } 
{\prod_{l=1}^{f_q+1}\a_p-\a_q -r^{(p)}_{k-1} -s^{(q)}_l +r^{(q)}_{l-1} +s^{(p)}_k}
\bigg \}\,,\\        
\frac
{G_{Y_p, Y_p}(0)}{G_{Y^{(k,+)}_p, Y^{(k,+)}_p}(0)}
&=& 
\frac {\prod_{l=1}^{f_p} -r^{(p)}_{k-1} -s^{(p)}_l +r^{(p)}_l +s^{(p)}_k}
{\prod_{l\neq k}^{f_p +1}-r^{(p)}_{k-1} -s^{(p)}_{l} +r^{(p)}_{l-1} +s^{(p)}_k}\,.
\ea
We substitute the above three equations to (\ref{Q+a}), we obtain (\ref{Q+}).
The derivation of (\ref{Q-}) from (\ref{Q-a}) is similar.


\section{Proof of Eq.(\ref{xy})}
\label{s:prcomb}
Let $x_I$ ($I=1,\cdots, N$) be arbitrary complex numbers.
We first observe,
\ba
\sum_{I=1}^N \prod_{J(\neq I)}^N \frac{1}{x_I-x_J}=0\,.
\ea
If we apply it to a set of variables $\{ x_1, \cdots, x_n, \xi \}$, ($\xi=x_{N+1}$) 
one derives,
\ba
&&\sum_{I=1}^N\frac{1}{\xi-x_I}\prod_{J (\neq I)}\frac{1}{x_I-x_J}=
-\sum_{I=1}^N \prod_{J(\neq I)}^{N+1} \frac{1}{x_I-x_J}
=\prod_{J=1}^N\frac{1}{\xi-x_J}
=:\frac{1}{\xi^N}\sum_{n=0}^\infty
\frac{b_n(x)}{\xi^{n}} \, .
\ea
The function $b_n(x)$ defined in the last line can be written as,
\ba
b_n(x)=\sum_{I_1\leq\cdots \leq I_n }x_{I_1}\cdots x_{I_n}\,.
\ea
The first part of this equality can be expanded as,
$\sum_{n=0}^\infty\sum_I \frac{(x_I)^n}{\xi^{n+1}}\prod_{J (\neq I)}\frac{1}{x_I-x_J}$.  So we derived,
\ba
\sum_{I=1}^N (x_I)^n\prod_{J (\neq I)}^N \frac{1}{x_I-x_J}=\left\{
\begin{array}{ll}
0\quad & n=0,\cdots, N-2\\
b_{n-N+1}(x) \quad& n\geq N-1
\end{array}
\right.\,.
\ea
If we write $\prod_{I=1}^M (\xi+y_J)=\sum_{n=0}^M \xi^n f_{M-n}(y)$ with
\ba
f_n(x)=\sum_{I_1<\cdots<I_n} x_{I_1}\cdots x_{I_n} \, ,
\ea
the left hand side of (\ref{xy}) is written as,
\ba
\sum_{I=1}^N \frac{\prod_{I=1}^M (x_I+y_J)}{\prod_{J (\neq I)}^N (x_I-x_J)}
=\sum_{n=0}^M f_{M-n}(y) \sum_{I=1}^N \frac{(x_I)^n}{\prod_{J (\neq I)}^N (x_I-x_J)}
=\sum_{n=N-1}^M f_{M-n}(y) b_{n-N+1}(x)  \, .
\ea
It is not difficult to show that the last quantity is the coefficient of $\zeta^{M-N+1}$
of the function $\frac{\prod_{J=1}^\mathcal{M} (\zeta+y_J)}{\prod_{I=1}^\mathcal{N} (\zeta-x_I)}$.

\section{Proof of Virasoro constraint}\label{a:Virasoro}
\subsection{Proof for $L_{\pm 1}$}
The quantity to be evaluated is,
\ba
\sum_{I=1}^{\mathcal N}
x_I\frac{\prod_{J=1}^{\mathcal N}(x_I +y_J)}{\prod_{J \neq I}^{\mathcal N}(x_I -x_J)}
=\sum_{I<J}^{\mathcal N}y_I y_J +\sum_{I,J}^{\mathcal N}y_I x_J
+\sum_{I<J}^{\mathcal N}x_I x_J +\sum_{I}^{\mathcal N}x_I^2\,.
\ea
We rewrite it explicitly,
\ba
\sum_{I,J}^{\mathcal N}y_I x_J 
&=&\sum_{p}^{N}\sum_{q}^{N}\sum_{k}^{{f}_p}\sum_{l}^{f_q+1}-(a_p + \nu + s^{(p)}_k - r^{(p)}_{k})
(a_q + \nu + s^{(q)}_l - r^{(q)}_{l-1}) \nt
&&
+\sum_{p}^{N}\sum_{q}^{N}\sum_{k}^{{f}_p}\sum_{l}^{\bar{f}_q}-(a_p + \nu + s^{(p)}_k - r^{(p)}_{k})
(b_q + \nu -\mu + u^{(q)}_l - t^{(q)}_{l})  \nt
&&+\sum_{p}^{N}\sum_{q}^{N}\sum_{k}^{\bar{f}_p+1}\sum_{l}^{{f}_q+1}-(b_p + \nu -\mu + u^{(p)}_k - t^{(p)}_{k-1})
(a_q + \nu + s^{(q)}_l - r^{(q)}_{l-1})\nt
&&
+\sum_{p}^{N}\sum_{q}^{N}\sum_{k}^{\bar{f}_p+1}\sum_{l}^{\bar{f}_q}-(b_p + \nu -\mu + u^{(p)}_k - t^{(p)}_{k-1})
(b_q + \nu -\mu + u^{(q)}_l - t^{(q)}_{l})\,, \\
\sum_{I<J}^{\mathcal N}x_I x_J 
&=&\sum_{p<q}^{N}\sum_{k}^{f_p+1}\sum_{l}^{f_q+1}(a_p + \nu + s^{(p)}_k - r^{(p)}_{k-1})
(a_q + \nu + s^{(q)}_l - r^{(q)}_{l-1})\nt
&&
+
\sum_{p}^{N}\sum_{k<l}^{f_p+1}(a_p + \nu + s^{(p)}_k - r^{(p)}_{k-1}) 
(a_p + \nu + s^{(p)}_l - r^{(p)}_{l-1}) 
\nt
&&+\sum_{p}^{N}\sum_{q}^{N}\sum_{k}^{f_p+1}\sum_{l}^{\bar{f}_q}(a_p + \nu + s^{(p)}_k - r^{(p)}_{k-1})
(b_q + \nu -\mu + u^{(q)}_l - t^{(q)}_{l})  \nt
&&+\sum_{p<q}^{N}\sum_{k}^{\bar{f}_p}\sum_{l}^{\bar{f}_q}(b_p + \nu -\mu + u^{(p)}_k - t^{(p)}_{k})
(b_q + \nu -\mu + u^{(q)}_l - t^{(q)}_{l})\nt
&&
+
\sum_{p}^{N}\sum_{k<l}^{\bar{f}_p}(b_p + \nu -\mu + u^{(p)}_k - t^{(p)}_{k})
(b_p + \nu -\mu + u^{(p)}_l - t^{(p)}_{l})\,,\\
\sum_{I<J}^{\mathcal N}y_I y_J 
&=&\sum_{p<q}^{N}\sum_{k}^{f_p}\sum_{l}^{f_q}(a_p + \nu + s^{(p)}_k - r^{(p)}_{k})
(a_q + \nu + s^{(q)}_l - r^{(q)}_{l})
\nt
&&+
\sum_{p}^{N}\sum_{k<l}^{f_p}(a_p + \nu + s^{(p)}_k - r^{(p)}_{k}) 
(a_p + \nu + s^{(p)}_l - r^{(p)}_{l}) 
\nt
&&+\sum_{p}^{N}\sum_{q}^{N}\sum_{k}^{f_p}\sum_{l}^{\bar{f}_q+1}(a_p + \nu + s^{(p)}_k - r^{(p)}_{k})
(b_q + \nu -\mu + u^{(q)}_l - t^{(q)}_{l-1})  \nt
&&+\sum_{p<q}^{N}\sum_{k}^{\bar{f}_p+1}\sum_{l}^{\bar{f}_q+1}(b_p + \nu -\mu + u^{(p)}_k - t^{(p)}_{k-1})
(b_q + \nu -\mu + u^{(q)}_l - t^{(q)}_{l-1})\nt
&&
+
\sum_{p}^{N}\sum_{k<l}^{\bar{f}_p+1}(b_p + \nu -\mu + u^{(p)}_k - t^{(p)}_{k-1})
(b_p + \nu -\mu + u^{(p)}_l - t^{(p)}_{l-1})\,,\\
\sum_{I}^{\mathcal N}x_I^2
&=&\sum_{p}^{N}\sum_{k}^{f_p+1}(a_p + \nu + s^{(p)}_k - r^{(p)}_{k-1})^2
+
\sum_{p}^{N}\sum_{k}^{\bar{f}_p}(b_p + \nu -\mu + u^{(p)}_k - t^{(p)}_{k})^2\,.
\ea
Sum the above four equations together, we find most of the cross terms cancel with each other, and the remaining is 
\ba
\label{remain}
&&\sum_{p}^{N}(a_p + \nu)^2 + \sum_{p<q}^{N}(a_p + \nu)(a_q + \nu) -\sum_{p,q}^{N}(a_p + \nu)(b_q + \nu -\mu)\nt
&&  + \sum_{p<q}^{N}(b_p + \nu -\mu)(b_q + \nu -\mu)
 +\sum_{p}^{N}\sum_{k}^{f_p}  s^{(p)}_k (r^{(p)}_k - r^{(p)}_{k-1}) 
 -\sum_{p}^{N}\sum_{k}^{\bar{f}_p}  u^{(p)}_k (t^{(p)}_k - t^{(p)}_{k-1}) 
 \nt
&&=\frac{1}{2}|\vec{a}+\nu \vec{e}|^2-\frac{1}{2}|\vec{b}+(\nu-\mu)\vec{e}|^2+\frac{1}{2}\k^2+|\vec Y|-|\vec W| 
\, ,
\ea
where we have used
 $
-\sum_{p=1}^{N}(a_p-b_p+\mu)=\k 
$.

\subsection{Proof for $L_{\pm 2}$}

The strategy is almost the same as the $L_{\pm 1}$ case, but with the help of the following formula.
\begin{equation}
\begin{split}
\frac{G_{Y,Y}(0)}{G_{Y^{(k,+2E)},Y^{(k,+2E)}}(0)} 
&= 
\frac{1}{2} \times \frac{\prod_{l=1}^{f_i} (r_{k-1} +s_l -r_l -s_k)(r_{k-1} +s_l -r_l -s_k -1)}
{\prod_{l\neq k}^{f_i +1}(r_{k-1} +s_{l} -r_{l-1} -s_k)(r_{k-1} +s_{l} -r_{l-1} -s_k-1)}\,,
\end{split}
\end{equation}

\ba
\frac{g_{A^{(k,+2E)}, B}(x)}{g_{A, B}(x)}
&=&
 \frac{\prod_{l=1}^{f_B+1} (x +r^A_{k-1} +s^B_l -r^B_{l-1} -s^A_k)(x +r^A_{k-1} +s^B_l -r^B_{l-1} -s^A_k-1)}
{\prod_{l=1}^{f_B} (x +r^A_{k-1} +s^B_l -r^B_l -s^A_k) (x +r^A_{k-1} +s^B_l -r^B_l -s^A_k-1)} \,,  \\        
\frac{g_{A^{(k,+2H)}, B}(x)}{g_{A, B}(x)}
&=&
 \frac{\prod_{l=1}^{f_B+1} (x +r^A_{k-1} +s^B_l -r^B_{l-1} -s^A_k)(x +r^A_{k-1} +s^B_l -r^B_{l-1} -s^A_k+1)}
{\prod_{l=1}^{f_B} (x +r^A_{k-1} +s^B_l -r^B_l -s^A_k) (x +r^A_{k-1} +s^B_l -r^B_l -s^A_k+1)} \,.
\ea
Here we have made use of the following relations
\begin{equation}
\frac{[ x ]_{A^{(k,+2E)}}}{[ x ]_A}
=(x -r_{k-1} +s_k)(x -r_{k-1} +s_k +1)\,,
\quad
\frac{[ x ]_{A^{(k,+2H)}}}{[ x ]_A}
=(x -r_{k-1} +s_k)(x -r_{k-1} +s_k -1)\,.
\end{equation}

For the Young diagrams with arbitrary shape,
for the location where a vertical two-box cannot be added, it means $s_{k-1} = s_k +1$,
so we have 
\begin{equation}
A_{(p)k}+1
=a_p +\nu+s^{(p)}_k - r^{(p)}_{k-1} +1
=a_p +\nu+s^{(p)}_{k-1} - r^{(p)}_{k-1}
=B_{(p)k-1}\,,
\end{equation}
which leads to 
\begin{equation}
(A_{(p)k}-B_{(p)k-1})(A_{(p)k}-B_{(p)k-1}+1) = 0\,.
\end{equation}
This is a factor of the corresponding vertical box term, which makes it become zero.

Similarly, for the place where a horizontal two-box cannot be added, we have $r_{k} = r_{k-1} +1$ and $A_{(p)k}-1= B_{(p)k}$,
the corresponding horizontal term becomes zero.



\begin{thebibliography}{99}
\bibitem{r:Nekrasov}
 N.~A.~Nekrasov,
  ``Seiberg-Witten prepotential from instanton counting,''
  arXiv:hep-th/0306211;\\
N.~Nekrasov and A.~Okounkov,
  ``Seiberg-Witten theory and random partitions,''
  arXiv:hep-th/0306238.

\bibitem{r:AGT}
  L.~F.~Alday, D.~Gaiotto and Y.~Tachikawa,
  ``Liouville Correlation Functions from Four-dimensional Gauge Theories,''
  arXiv:0906.3219 [hep-th].


\bibitem{Nekrasov:2009rc} 
  N.~A.~Nekrasov and S.~L.~Shatashvili,
  ``Quantization of Integrable Systems and Four Dimensional Gauge Theories,''
  arXiv:0908.4052 [hep-th].

\bibitem{r:Virasoro}
  M.~Fukuma, H.~Kawai and R.~Nakayama,
  ``Continuum Schwinger-dyson Equations And Universal Structures In Two-dimensional Quantum Gravity,''
  Int.\ J.\ Mod.\ Phys.\ A {\bf 6}, 1385 (1991);\\
  R.~Dijkgraaf, H.~L.~Verlinde and E.~P.~Verlinde,
  ``Loop equations and Virasoro constraints in nonperturbative 2-D quantum gravity,''
  Nucl.\ Phys.\ B {\bf 348} (1991) 435;\\
  A.~Mironov and A.~Morozov,
  ``On the origin of Virasoro constraints in matrix models: Lagrangian approach,''
  Phys.\ Lett.\ B {\bf 252}, 47 (1990).

\bibitem{r:IM1}
  H.~Itoyama and Y.~Matsuo,
  ``Noncritical Virasoro algebra of $d < 1$ matrix model and quantized string field,''
  Phys.\ Lett.\ B {\bf 255}, 202 (1991).

\bibitem{r:W}
  M.~Fukuma, H.~Kawai and R.~Nakayama,
  ``Infinite dimensional Grassmannian structure of two-dimensional quantum gravity,''
  Commun.\ Math.\ Phys.\  {\bf 143}, 371 (1992).

\bibitem{r:IM2}
  H.~Itoyama and Y.~Matsuo,
  ``W(1+infinity) type constraints in matrix models at finite N,''
  Phys.\ Lett.\ B {\bf 262}, 233 (1991).
\bibitem{Dijkgraaf:2009pc} 
  R.~Dijkgraaf and C.~Vafa,
  ``Toda Theories, Matrix Models, Topological Strings, and N=2 Gauge Systems,''
  arXiv:0909.2453 [hep-th].

\bibitem{Alba:2009ya} 
  V.~Alba and A.~.Morozov,
  ``Check of AGT Relation for Conformal Blocks on Sphere,''
  Nucl.\ Phys.\ B {\bf 840}, 441 (2010)
  [arXiv:0912.2535 [hep-th]].

\bibitem{Wyllard:2009hg} 
  N.~Wyllard,
  ``A(N-1) conformal Toda field theory correlation functions from conformal N = 2 SU(N) quiver gauge theories,''
  JHEP {\bf 0911}, 002 (2009)
  [arXiv:0907.2189 [hep-th]].

\bibitem{Mironov:2009by} 
  A.~Mironov and A.~Morozov,
  ``On AGT relation in the case of U(3),''
  Nucl.\ Phys.\ B {\bf 825}, 1 (2010)
  [arXiv:0908.2569 [hep-th]].

\bibitem{Alba:2010qc} 
  V.~A.~Alba, V.~A.~Fateev, A.~V.~Litvinov and G.~M.~Tarnopolskiy,
  ``On combinatorial expansion of the conformal blocks arising from AGT conjecture,''
  Lett.\ Math.\ Phys.\  {\bf 98}, 33 (2011)
  [arXiv:1012.1312 [hep-th]].

\bibitem{Fateev:2011hq} 
  V.~A.~Fateev and A.~V.~Litvinov,
  ``Integrable structure, W-symmetry and AGT relation,''
  JHEP {\bf 1201}, 051 (2012)
  [arXiv:1109.4042 [hep-th]].

\bibitem{Belavin:2011js} 
  A.~Belavin and V.~Belavin,
  ``AGT conjecture and Integrable structure of Conformal field theory for c=1,''
  Nucl.\ Phys.\ B {\bf 850}, 199 (2011)
  [arXiv:1102.0343 [hep-th]].

\bibitem{Kanno:2011qv} 
  S.~Kanno, Y.~Matsuo and S.~Shiba,
  ``W(1+infinity) algebra as a symmetry behind AGT relation,''
  Phys.\ Rev.\ D {\bf 84}, 026007 (2011)
  [arXiv:1105.1667 [hep-th]].

\bibitem{Estienne:2011qk} 
  B.~Estienne, V.~Pasquier, R.~Santachiara and D.~Serban,
  ``Conformal blocks in Virasoro and W theories: Duality and the Calogero-Sutherland model,''
  Nucl.\ Phys.\ B {\bf 860}, 377 (2012)
  [arXiv:1110.1101 [hep-th]].

\bibitem{Schomerus:2003vv} 
  V.~Schomerus,
  ``Rolling tachyons from Liouville theory,''
  JHEP {\bf 0311}, 043 (2003)
  [hep-th/0306026].

\bibitem{Awata:1994tf} 
  E.~Frenkel, V.~Kac, A.~Radul and W.~-Q.~Wang,
  ``W(1+infinity) and W(gl(N)) with central charge N,''
  Commun.\ Math.\ Phys.\  {\bf 170}, 337 (1995)
  [hep-th/9405121];\\
  H.~Awata, M.~Fukuma, Y.~Matsuo and S.~Odake,
  ``Representation theory of the W(1+infinity) algebra,''
  Prog.\ Theor.\ Phys.\ Suppl.\  {\bf 118}, 343 (1995)
  [hep-th/9408158].

\bibitem{Mironov:2010pi} 
  A.~Mironov, A.~Morozov and S.~.Shakirov,
  ``A direct proof of AGT conjecture at beta = 1,''
  JHEP {\bf 1102}, 067 (2011)
  [arXiv:1012.3137 [hep-th]].

\bibitem{Zhang:2011au} 
  H.~Zhang and Y.~Matsuo,
  ``Selberg Integral and SU(N) AGT Conjecture,''
  JHEP {\bf 1112}, 106 (2011)
  [arXiv:1110.5255 [hep-th]].

\bibitem{Itoyama:2010ki} 
  H.~Itoyama and T.~Oota,
  ``Method of Generating q-Expansion Coefficients for Conformal Block and N=2 Nekrasov Function by beta-Deformed Matrix Model,''
  Nucl.\ Phys.\ B {\bf 838}, 298 (2010)
  [arXiv:1003.2929 [hep-th]].



\bibitem{Dotsenko:1984nm} 
  V.~S.~Dotsenko and V.~A.~Fateev,
  ``Conformal Algebra and Multipoint Correlation Functions in Two-Dimensional Statistical Models,''
  Nucl.\ Phys.\ B {\bf 240}, 312 (1984).

\bibitem{r:Sato}
E. Date, M. Jimbo, M.Kashiwara and T. Miwa, "Transformation groups
for soliton equations I--VI", Proc. Japan Acad. 57A, 342, 387 (1981);
J. Phys. Soc. Japan 50, 3806, 3813 (1982); Physica 4D, 343 (1982);
Publ. RIMS. 18, 1077, 1111 (1982);\\
Mikio Sato, "Soliton equation and Universal Grassmannian manifold",
(note taken by M. Noumi), Jochi University, 1984.

\bibitem{r:KMS}
 S.~Kanno, Y.~Matsuo and S.~Shiba,
  ``Analysis of correlation functions in Toda theory and AGT-W relation for SU(3) quiver,''
  Phys.\ Rev.\ D {\bf 82}, 066009 (2010)
  [arXiv:1007.0601 [hep-th]].
\bibitem{r:DP}
  N.~Drukker and F.~Passerini,
  ``(de)Tails of Toda CFT,''
  JHEP {\bf 1104}, 106 (2011)
  [arXiv:1012.1352 [hep-th]].
\bibitem{r:Winf}
For example,\\
  S.~Iso, D.~Karabali and B.~Sakita,
  ``Fermions in the lowest Landau level: Bosonization, W infinity algebra, droplets, chiral bosons,''
  Phys.\ Lett.\ B {\bf 296}, 143 (1992)
  [hep-th/9209003]:\\
  A.~Cappelli, C.~A.~Trugenberger and G.~R.~Zemba,
  ``Stable hierarchical quantum hall fluids as W(1+infinity) minimal models,''
  Nucl.\ Phys.\ B {\bf 448}, 470 (1995)
  [hep-th/9502021];\\
  M.~Henneaux and S.~-J.~Rey,
  ``Nonlinear $W_{infinity}$ as Asymptotic Symmetry of Three-Dimensional Higher Spin Anti-de Sitter Gravity,''
  JHEP {\bf 1012}, 007 (2010)
  [arXiv:1008.4579 [hep-th]];\\
  M.~R.~Gaberdiel, R.~Gopakumar and A.~Saha,
  ``Quantum $W$-symmetry in $AdS_3$,''
  JHEP {\bf 1102}, 004 (2011)
  [arXiv:1009.6087 [hep-th]].

\bibitem{r:SV}
Olivier Schiffmann, Eric Vasserot, 
``Cherednik algebras, W algebras and the equivariant 
cohomology of the moduli space of instantons on $A^2$'',
arXiv:1202.2756.

\end{thebibliography}
\end{document}